# High temperature plasmonics: Narrowband, tunable, near-field thermal sources

Yu Guo, S. Molesky, C. Cortes and Zubin Jacob[*]

*Department of Electrical and Computer Engineering,*

*University of Alberta, Edmonton, Alberta T6G 2V4, Canada*

[*]*zjacob@ualberta.ca*

**Abstract:** We propose multiple approaches for controlling Wien's displacement law in the near-field leading to narrowband, tunable, spatially-coherent high temperature thermal sources. We show that narrowband super-planckian thermal emission relevant to the wavelength ranges for near-field thermophotovoltaics can be achieved by utilizing engineered plasmonic states of metals with a high melting point. Along with conventional thin film surface plasmon polaritons of high temperature plasmonic metals, we also introduce multiple different approaches to narrowband near field thermal emission a) epsilon-near-zero metamaterials b) tunable surface-plasmon-polaritons of anisotropic metamaterials and c) photonic van hove singularities. Our work solves the critical issue of a narrowband thermal source for near-field thermophotovoltaics.

Nanoengineering the coherent state of light as in a SPASER [1] and the quantum state of light as in nanoscale single photon sources [2] have received significant attention over the last few years. However, controlling the thermal state of light i.e. nanoscale thermal sources with spectral tuning, narrow bandwidth and spatial coherence remains a challenge. In this paper, we show that high temperature plasmonic metamaterials form an ideal platform for controlling thermally excited radiation in the near field. We note that 3D fabrication tolerances pose a severe limitation to narrowband operation of tungsten photonic crystals for thermal engineering. On the other hand, surface polariton approaches for tailoring near-field thermal emission which rely on material resonances (eg: SiC in the mid-IR) and are not tunable [3]. This is especially important for applications such as thermophotovoltaics where the thermal emission has to be between 1 μm < λ < 2 μm (T ≈ 1500 K) [4].

Our approach utilizes three design principles i) a switch away from conventional plasmonic building blocks like silver and gold to alternate plasmonic materials based on oxides and nitrides [4,5]. We note that they have high melting points in the range of 3000 C allowing operation in the near-IR range crucial for practical applications; ii) engineering the epsilon-near-zero resonances and surface plasmon polaritons using anisotropic multilayer plasmonic metamaterials. iii) We also introduce the concept of a photonic van hove singularities (VHS) in metal-dielectric structures that arise due to slow light modes [6] for tunable narrowband near-field thermal emission (NFTE). (Fig. 1 and 2) We also consider the potential application in near-field thermophotovoltaics where the narrowband thermal emission has to be 1 μm < λ < 2 μm (T ≈ 1500 K) for efficient energy conversion beyond the Shockley Queisser limit using low bandgap GaSb photovoltaic cells. (Fig. 2 c) We note that using thin films of Titanium Nitride or Aluminum Zinc Oxide should lead to surface plasmon polaritons in the near-infrared range critical for thermophotovoltaics.

In figure 1 (a), we consider a multilayer metal-dielectric structure consisting of aluminum zinc oxide (AZO) and zinc oxide (ZnO). They achieve an effectively anisotropic response. The thermal excitation of modes of this structure leads to a narrowband peak in the near field (calculated using fluctuational electrodynamics). To understand the origin of the peak, we plot the wavevector resolved local density of states for the structure in fig 1 (b). The bright bands correspond to the modes of the structure which are thermally excited. It is seen that the peak corresponds to the wavelength where the group velocity of the mode approaches zero. We call this a photonic van hove singularity that arises due to slow light modes in this coupled-plasmonic structure. In figure 1c we analyze the near-field spatial coherence [7] of our narrowband tunable thermal source. We see an excellent spatial coherence due to the preferential thermal excitation of the slow light mode.

We now show another route of achieving a narrowband super-planckian thermal source. Multilayer metal-dielectric metamaterials also support epsilon-near-zero resonances and tunable anisotropic surface plasmon polariton resonances along with hyperbolic modes [8,9]. Hyperbolic dispersion leads to broadband super-planckian thermal emission. When the metal fill fraction is varied, we notice the narrow peak of thermal emission due to ENZ and anisotropic SPP modes to be spectrally tuned (fig 2 a and b).

Finally, we show that the near field heat transfer [10–12] for our multilayer structures supporting the van hove singularity can far exceed the value in tungsten as well as surface-plasmon-polaritons of bulk media. In fig. 2(c), we plot the heat transfer between multilayer AZO/ZnO structures and compare it to tungsten as well as bulk AZO slab. The temperature of the two slabs are taken to be T1=1500 K and T2 = 300 K. First we note the excellent agreement between effective medium predictions (EMT) of the location of the van hove singularity and the practical multilayer structure (Transfer matrix method - TMM). Second, we note that the van hove singularity outperforms both tungsten as well as surface plasmon polaritons of bulk AZO slab.

In conclusion, we have shown the potential of high temperature plasmonic metamaterials as narrowband tunable thermal sources for applications in thermophotovoltaics and near-field thermal stamping. Our work also paves the way for manipulating the near-field spatial coherence and bandwidth of thermal sources using metamaterials.

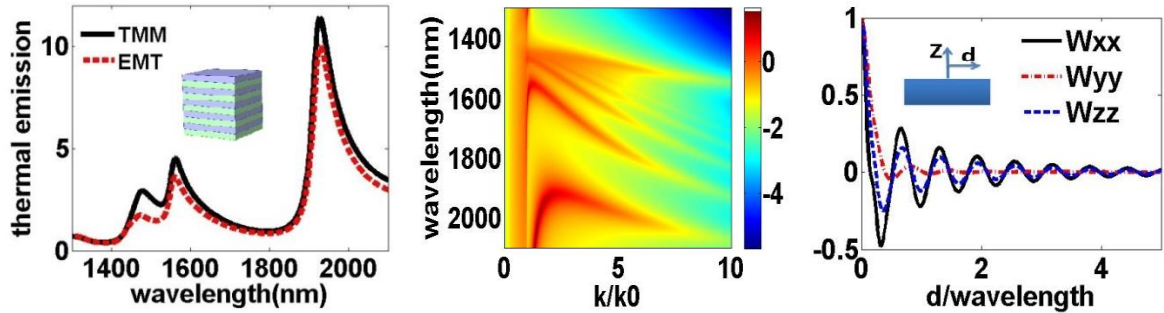

**Fig 1**: (left) Narrowband tunable near-field thermal peaks calculated using fluctuational electrodynamics (normalized to the black body radiation into the upper half-space). The wavelength range between 1 μm and 2 μm is ideally suited for thermophotovoltaic applications due to the existence of low bandgap TPV cells such as Gallium Antimonide (0.7 ev). The peaks occur due to slow light modes in multilayer structures consisting of Aluminum doped zinc oxide and Zinc oxide. These can be considered as photonic van-hove singularities where the density of states and thermal emission is enhanced. Note these plasmonic materials have a high melting point (~ 3000 C) as compared to silver and gold (~ 1000 C) paving the way for high temperature plasmonics. We have taken into account loss, dispersion, finite unit cell and finite sample size in our calculations. (Parameters: AZO and ZnO ($\varepsilon=6$) multilayers, each layer 25nm and 20 layers in total, the distance is 150nm.) (center) Wave-vector resolved thermal emission. The bright bands denote the modes of the structure which are thermally excited. We can clearly see a slow light mode around 1900nm where the slope of the band goes to zero. (right) Spatial coherence at the wavelength (1927nm) which supports the slow light mode. Note that the thermal excitation of the slow light mode (coupled-plasmonic state) gives rise to spatial coherence in the near field.

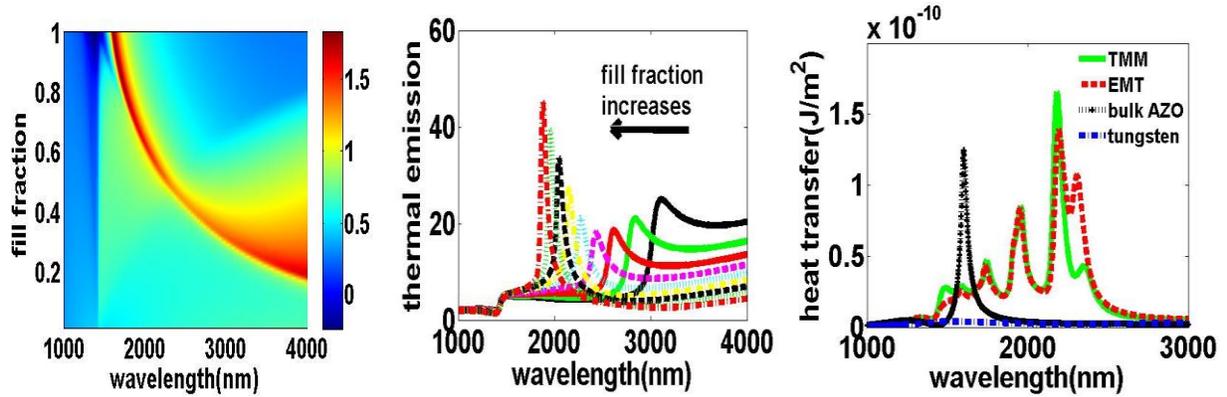

**Fig 2**: (left) Near field thermal emission (log scale plot) due to epsilon-near-zero resonances and anisotropic surface plasmon polaritons in multilayer metamaterials. The bright red band is the super-planckian thermal emission which can be tuned by changing the fill fraction. (center) Thermal emission spectrum for fill fractions varying from 0.3 to 0.7. Clearly, we can tune the thermal emission peaks within a wide spectrum. (right) Heat transfer using AZO and ZnO ($\varepsilon=6$) multilayered structure with a gap distance 200nm. The photonic van hove singularity arising due to coupled plasmons of the multilayer system has multiple heat transfer peaks which can beat tungsten and even the surface plasmon polaritons of bulk AZO. This implies high temperature plasmonic metamaterials can be used for near-field thermophotovoltaics and other near-field applications with better performance than bulk metals or widely used tungsten. Parameters: (left) and (center) semi-infinite layers calculated with EMT, at a distance 100nm. (right) AZO and ZnO multilayers, each layer 25nm and 50 layers in total, the distance is 200nm, Heat transfer parameters $T_1= 1500$ K, $T_2 = 300$ K.